# Beyond Bean's critical state model: On the origin of paramagnetic Meissner effect

**Sangjun Oh**[1†], **Dong Keun Oh**[1], **Won Nam Kang**[2], **Jung Ho Kim**[3], **Shi Xue Dou**[3], **Dojun Youm**[4] **and Dong Ho Kim**[5]

[1]National Fusion Research Institute, Daejeon 305-333, Korea.
[2]Department of Physics, Sungkyunkwan University, Suwon 440-746, Korea.
[3]University of Wollongong, Northfields Avenue, Wollongong, NSW 2522, Australia.
[4]Department of Physics, Korea Advanced Institute of Science and Technology, Daejeon 305-701, Korea.
[5]Department of Physics, Yeungnam University, Gyeongsan 712-749, Korea.

[†]Correspondence to: wangpi@nfri.re.kr

**Abstract.** Solving phenomenological macroscopic equations instead of microscopic Ginzburg-Landau equations for superconductors is much easier and can be advantageous in a variety of applications. However, till now, only Bean's critical state model is available for the description of irreversible properties. Here we propose a plausible overall macroscopic model for both reversible and irreversible properties, combining London theory and Bean's model together based on superposition principle. First, a simple case where there is no pinning is discussed, from which a microscopic basis for Bean's model is explored. It is shown that a new concept of 'flux share' is needed when the field is increased above the lower critical field. A portion of magnetic flux is completely shielded, named as 'Meissner share' and the rest penetrates through vortices, named as 'vortices share'. We argue that the flux shares are irreversible if there is pinning. It is shown that the irreversible flux shares can be the reason for observed peculiar reversible magnetization behavior near zero field. The overall macroscopic model seems to be valuable for the analysis of fundamental physical properties as well. As an example, it is shown the origin of paramagnetic Meissner effect can be explained by the phenomenological macroscopic model.

Superposition principle is implicitly assumed for the analysis of type II superconductors. Fundamental physical parameters, such as the Ginzburg-Landau parameter, can be extracted from reversible magnetization with an assumption that intrinsic superconducting properties are not affected by extrinsic pinning properties [1]. Extrinsic parameters, for example, the critical current density, can be calculated solely from irreversible magnetization by using Bean's critical state model [2]. However, in some cases, it is being questioned whether reversible and irreversible magnetizations are separable quantities. Superconducting volume fraction can be obtained from magnetization slope (-$dM/dH$) below the lower critical field. The slope is close to $4\pi$ for high quality samples when it is measured from initial magnetization loops [3], but usually pretty much reduced if estimated from reversible magnetization. Reversible magnetization is a mean between increasing and decreasing field branches. For both branches, vortices are trapped inside and therefore it seems to be natural to claim that the field produced by the trapped vortices is the reason for that reduction [3]. The other interesting possibility has been argued by Hellman *et al*. The Meissner screening current may be interfered by irreversible current flow over a penetration depth [3]. At least for some cases, we need to consider the possibility of interaction between the two.

Reversible magnetization can be calculated from a thermodynamic analysis based on the Ginzburg-Landau theory. The above possible interaction may have been studied by solving the Ginzburg-Landau equations in detail, including the effect of pinning. Recently, Chapman has reported hierarchical models leading the Ginzburg-Landau theory down to empirical Bean's critical state model [4]. A bit unfortunately, to study the effect of pinning, a number of vortices need to be included and even more numbers of mesh points are required to solve the equations numerically, which is in fact impractical. On the other hand, in the calculation of irreversible magnetization by using Bean's model, each vortex does not need to be separately treated. Continuum electromagnetics can be adopted with an additional equation on the critical current density. Any complicated geometry can be easily simulated by using a commercial finite element method (FEM) software [5], very useful for practical applications. However, Bean's critical state model only can describe irreversible properties and therefore, to study the possible interaction we need to add another macroscopic model for reversible properties. Below the lower critical field, the Meissner effect can be easily described by a local theory, by the London equations. Hereafter, we will refer the macroscopic model based on London theory as 'London model'.

Putting the London and Bean's critical state models together, on the other hand, is also not a trivial problem. Even if we apply the superposition principle, we still need to know how we can separate reversible and irreversible parts. For instance, above the lower critical field, how much portion of external field needs to be solved by the London model and by Bean's critical state model? A bit more delicate problem is on the boundary. As was discussed by Hellman *et al*., do we need to think about possible interference between the Meissner screening current and irreversible current flow within a penetration depth? Or do we need to consider the field produced by trapped vortices on reversible magnetization? To address these problems, first, microscopic basis of Bean's critical state model needs to be discussed, to clarify whether Bean's model is suitable for this kind of analysis or not. In a zero critical current density limit, irreversible magnetization vanishes and the Bean's model reduces to a trivial solution, magnetic field is uniformly distributed throughout the sample. In the first section of this work, microscopic basis of Bean's critical state model will be discussed for this simple case when there is no pinning. In the second part, the argument will be extended to a pinned case. Finally, it will be discuss that the well-known paramagnetic Meissner effect [6] can be naturally understood by using the phenomenological macroscopic model for type II superconductors.

**Bean's critical state model from a microscopic point of view**

The main feature of Bean's critical state model is that dissipationless current can exist far inside a superconductor, which can be written as an empirical power law relation, $E = E_0(J/J_c)^n$, where $E_0$, $J_c$ are the electric field criterion and the critical current density, respectively. When it is combined with Faraday's law ($\nabla \times \boldsymbol{E} = -\mu_0 \, \partial \boldsymbol{H}/\partial t$) together with Ampere's law ($\boldsymbol{J} = \nabla \times \boldsymbol{H}$), a set of partial differential equations on $\boldsymbol{H}$ can be obtained, which can be solved by using a partial differential equation (PDE) mode of commercial FEM software [5]. As was emphasized by Brandt, the only way extrinsic current can exist inside a type II superconductor is when it is accompanied by the density gradient or curvature of vortices [7]. We first consider a simple case where there is no pinning, which corresponds to a zero critical current density limit of Bean's critical state model. An infinitely long superconducting beam with a square cross-section is schematically drawn in Fig. 1. Below the lower critical field, only Meissner shielding current will flow on the surface, which can be described by London theory and will be referred as 'London model' in this work. The London equations also can be easily incorporated into commercial FEM software. For example, if we choose the London gauge, the supercurrent density, $J_s$, is just proportional to the vector potential, $A$, $J_S = -n_S e^2 A/mc = -c/4\pi\lambda^2 \cdot A$, where $n_s$ is the number density of superconducting electrons, $\lambda$ is the penetration depth. As the field is increased above the lower critical field, vortices will be formed from the surface, distributed uniformly inside the superconductor so that there is no vortex density gradient, no extrinsic current, as shown in Fig. 1.

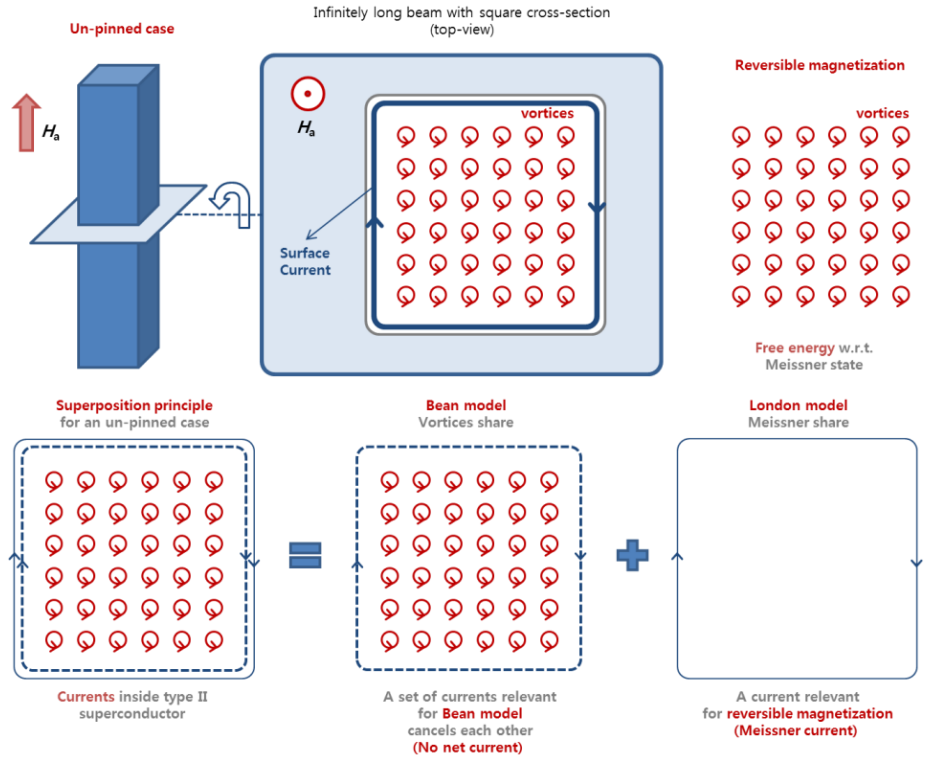

**Figure 1 | A schematic illustration for an infinitely long superconductor.** Sets of currents relevant for the Bean critical state model and London model are shown together, which are decomposed based on superposition principle.

In this simple geometry, a solution of the Ginzburg-Landau equations for an isolated vortex when the Ginzburg-Landau parameter, $\kappa$, far greater than 1 ($\kappa \gg 1$) can be easily found. The field profile is proportional to a zeroth-order Hankel function of imaginary argument [8]. (This solution sometimes also referred as 'London model'. But in this work, as noted, we refer the London model as a macroscopic model for the Meissner shielding current on the surface.) For arbitrary $\kappa$, a bit more realistic solution near vortex core can be obtained by using a variational method, by Clem model [9]. According to the Clem model [9], the magnetic field by each individual vortex, $b_z$, can be written as [9,10],

$$b_z \propto f_\infty K_0(f_\infty(\rho^2 + \xi_v^2)^{1/2})/\kappa\xi_v K_1(f_\infty\xi_v),$$

where $K_n(x)$ is a modified Bessel function of $n$th order. The two variational parameters, $\xi_v$ and $f_\infty$ are related with the effective core radius and the order parameter reduction by the effect of vortices overlap.

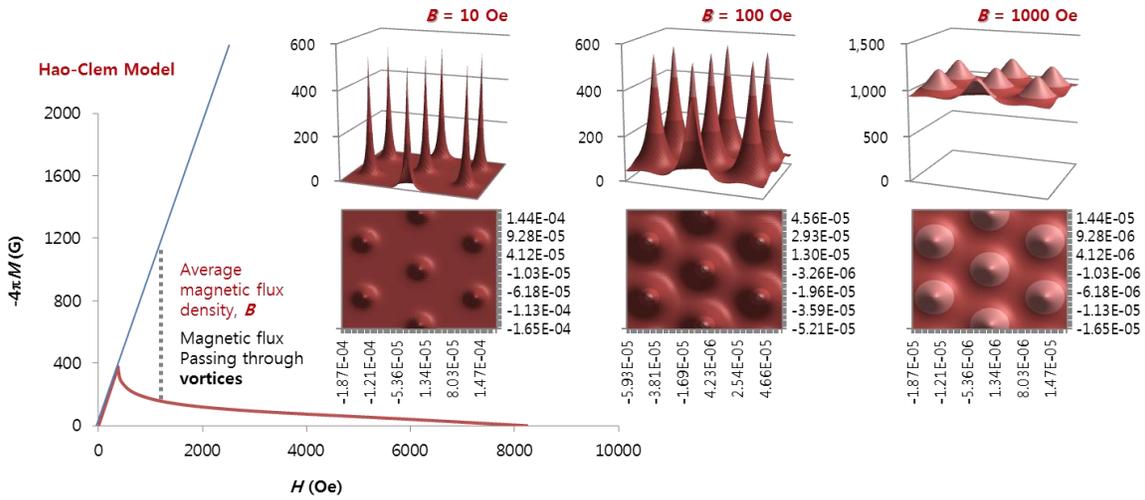

**Figure 2 | A typical reversible magnetization calculation result and field profiles inside a superconductor at some specific fields.** The reversible magnetization and the field profile by a vortex are calculated by Hao-Clem model. The Ginzburg-Landau parameter, $\kappa$, of 5 and the penetration depth, $\lambda$, of 100 nm are assumed.

These analytic solutions of the Ginzburg-Landau equations are quite useful for reversible magnetization calculation. A reversible magnetization calculation result by using Hao-Clem model [10], for the Ginzburg-Landau parameter of 5, is presented in Fig 2. The penetration depth, $\lambda$, is assumed as 100 nm. In the inset, field profiles for uniformly distributed hexagonal arrays of vortices are presented together. Here again, superposition principle can be applied so that the overall field profile is simply a sum of the field produced by each individual vortex.

Typically, for the calculation of reversible magnetization, only the vortices far away from the surface are considered so that the effect of surface current can be ignored [10]. The Ginzburg-Landau free energy is calculated with respect to a Meissner state, a state where there is no vortex and only the surface perfect diamagnetic current exists. Its derivative with respect to the magnetic flux density determines the thermodynamic magnetic field, $H$, and thereby the magnetization, $4\pi M$. As the magnetization is given by the derivative of the free energy, the free energy contribution by the surface current, which remains the same as the field is increased, can be ignored. Another usual assumption is on the average magnetic flux density inside a superconductor, $B$, assumed as proportional to vortex density [11], as a ratio, quantum flux divided by vortices lattice unit cell area. This uniform magnetic flux inside a superconductor when there is no pinning is what we can expect from Bean's critical state model.

Till now, it was discussed that for the simple un-pinned case, as schematically presented in Fig. 1, all the currents inside can be calculated by microscopic theory and far away from the surface, vortex array can be treated as uniform magnetic flux, compatible with microscopic theory on reversible magnetization. How all these currents inside can be grouped together into a macroscopic model? One possible idea is that a part of surface current constitutes Bean's critical state model together with vortices and the rest, solely responsible for reversible magnetization, as schematically presented in Fig. 1. This idea seems to be reasonable for two reasons. Firstly, only a part of surface current is responsible for reversible magnetization seems to be reasonable as the reversible magnetization decreases as the field is increased as shown in Fig. 2. As the field is increased less surface current must be involved in for reversible magnetization. The rest of surface current, relevant for Bean's critical state model, also seems to be necessary to balance the overall Lorentz force. More surface current relevant for Bean's model will be needed as the number of vortices increases, as the field is increased.

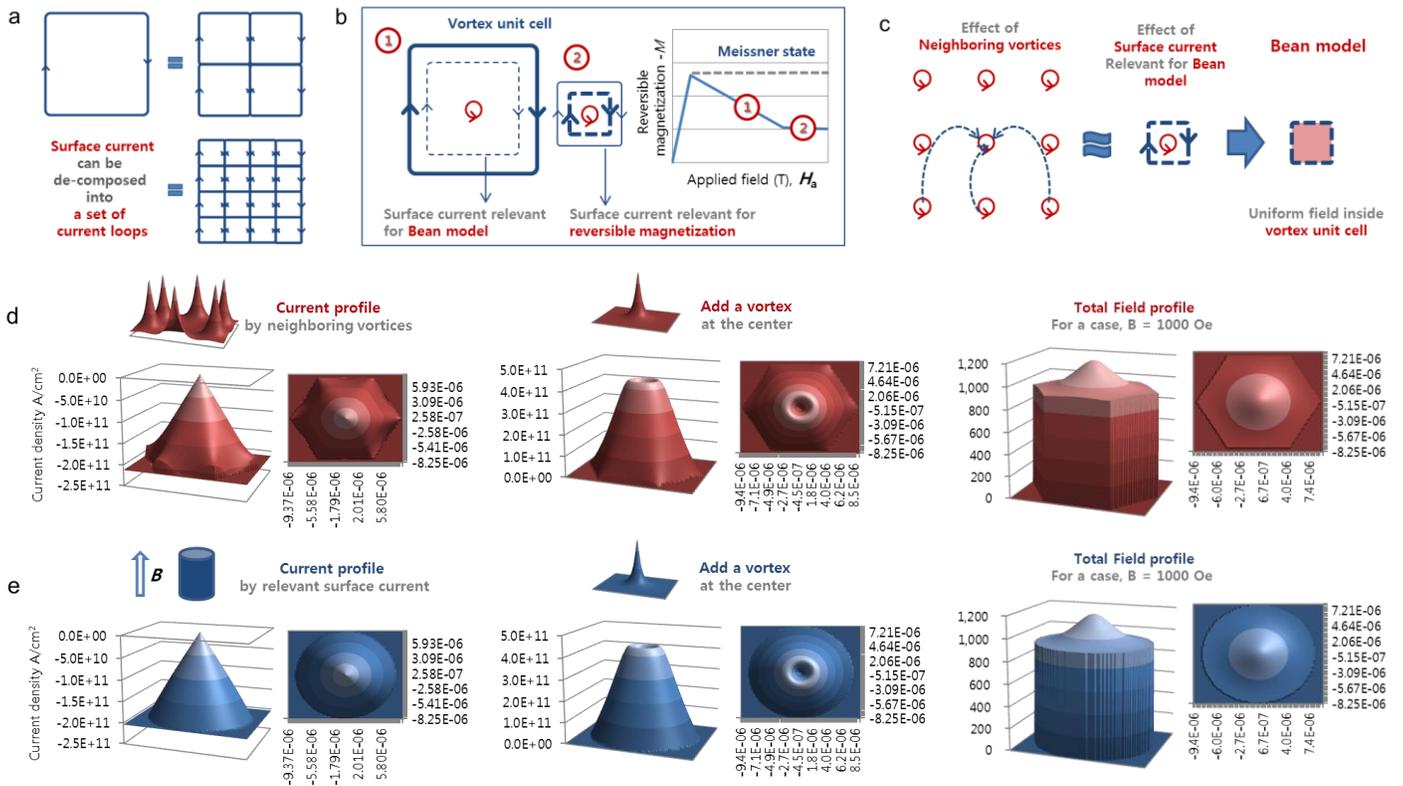

**Figure 3 | A microscopic building block for Bean's critical state model. a**, Surface current can be decomposed into a set of current loops. **b**, A vortex unit cell composed of a central vortex and surrounding reduced surface currents. The size of a vortex unit cell decreases as the field increases. **c**, The effect of all neighboring vortices can be equivalently represented by a reduced surface current relevant for Bean's model. **d**, From left to right, current profile by neighboring vortices, total current and field profiles including the central vortex within a vortex unit cell for ***B*** = 1000 Oe. **e**, From left to right, equivalent surface current profile, total current and field profiles including the central vortex within a relevant circular unit cell.

Is this picture on the surface current compatible with microscopic theory? Our strategy is schematically presented in Fig. 3. In the reversible magnetization calculation, the free energy within any vortex unit cell is equal to each other, due to the translational symmetry of vortex lattice. Likewise, the surface current will be subdivided down to the size of vortex unit cell as shown in Fig. 3a. As the overlapped parts cancel each other, the surface current can be divided into a set of current loops. Can this subdivided surface current further be decomposed into a part correspond to the London model and the rest for Bean's critical state model? In Fig. 3b, the field dependence of reversible magnetization is simplified as linearized lines. (The reversible magnetization by the Meissner state is constant as shown in Fig. 3b as a dotted line.) This decomposition is possible, if the decomposed surface current relevant for Bean's critical state model is microscopically equivalent to the overlapped currents by the neighboring vortices, as schematically depicted in Fig. 3c. Then it can be argued that each unit cell consists of the decomposed surface Bean current and a vortex at the center, which can be further simplified as a uniform magnetic field and can be regarded as a unit cell for Bean's critical state model.

In schematic figures, Fig. 1 and 3, square vortex lattices are depicted because of their easiness to draw. On the other hand, in Fig. 2, for the field profile calculations, hexagonal lattice is assumed as it is widely observed. For a hexagonal unit cell, the surface current can be reasonable approximated as a circular current. For an infinitely long cylinder of radius, $a$, analytic solution of the London equations can be found, which is described by another modified Bessel function. The field decay near the surface can be written as [12],

$$b_z = b_0 I_0(\rho/\lambda)/I_0(a/\lambda)$$

where $I_n(x)$ is another modified Bessel function of $n$th order and $b_0$ is the external field. One of the interesting features of the above solution is that the field is not zero at the center. The field at the center is almost zero when the radius, $a$, is much larger than the penetration depth, $\lambda$, but is increased and approaches the value of external field as the radius is reduced down to $\lambda$ and below. In Fig. 3d and 3e, the effect of neighboring vortices overlap and that of the decomposed surface Bean current is compared. The magnetic flux density, $B$ of 1000 Oe case is shown in detail, especially within the vortex unit cell. At 1000 Oe, the size of the unit cell is only about 89 nm, less than the penetration depth, $\lambda$ of 100 nm, so that the overlap effect is quite severe. The overlapped current flows clockwise as shown in Fig. 3d, the left side, of which maximum is about $2 \times 10^{11}$ A/cm$^2$. If we add a vortex at the center, the current at the unit cell boundary is almost cancelled to zero and the maximum counterclockwise current of about $4 \times 10^{11}$ A/cm$^2$ can be observed near the vortex core. The effective core radius, $\xi_v$, is about 26 nm. Overall field profile is shown in Fig. 3d, the right side. As expected, field overlap by neighboring vortices is quite high so that the average magnetic flux density within the unit cell is equal to 1000 Oe. The same can be argued for the decomposed surface Bean current as shown in Fig. 3e. For the calculation shown in Fig. 3e, external magnetic flux density, $b_0$ of 1000 Oe has been used, the same as the corresponding average flux density by the vortices. The same can be argued for other fields as well. (As a further example, $B$ = 10, and 100 Oe cases are presented in Supplementary Information.)

This equivalence between the effect of neighboring vortices and that of decomposed surface Bean current means that surface or boundary does not need to be specially treated. From this microscopic point of view, it may be argued that the Meissner screening current does not affect irreversible current flow over a penetration depth. Only a part of the Meissner screening current is relevant for Bean's critical state model and it is separable from the rest. Also, the effect of field produced by vortices is negligible as the vortex unit cell behaves basically the same way whether it is located near the surface or not. Now, it can be argued that the currents inside a superconductor indeed can be grouped together into the London and Bean's model as shown in Fig. 1. Each vortex unit cell can be further simplified as a uniform magnetic flux unit, as shown in Fig. 4a and this unit cell can be regarded as a sort of a microscopic building block for Bean's critical state model. The meaning of this separation can be more clearly seen from the field profiles as shown together in Fig. 4a. A part of flux just penetrates through vortices and the rest completely shield. In the above case shown in Fig. 3d, $B$ = 1000 Oe case, the calculated $H$ by Hao-Clem model is 1160 Oe. A portion of external magnetic field, 1000 Oe, out of total 1160 Oe can be separately described by Bean's critical state model as uniformly penetrating magnetic flux and the rest, 160 Oe, completely shielded, which can be described by the London model. In an analogical way, we may say that a superconductor in magnetic flux behaves in a similar way to a pebble in a stream, as schematically shown in Fig. 4b.

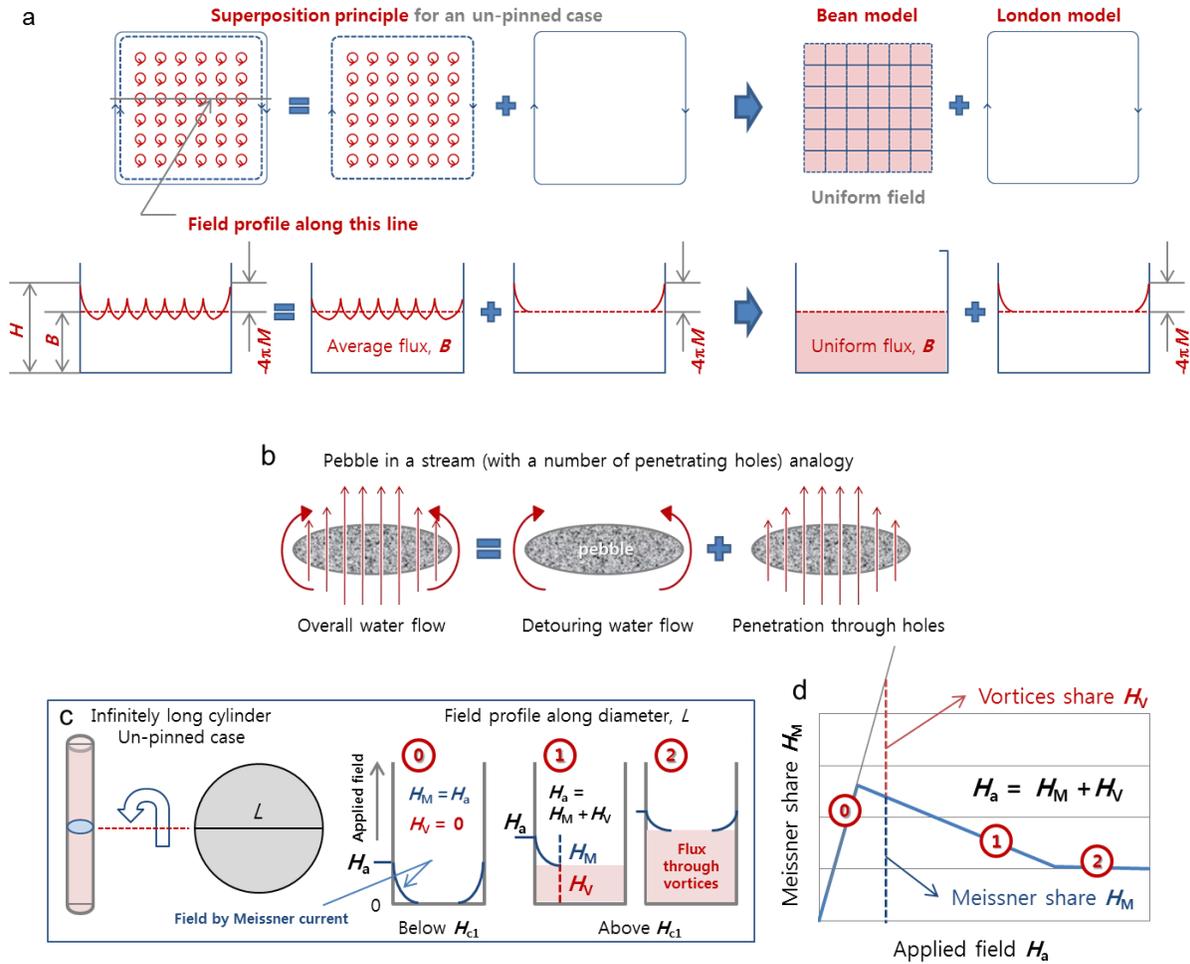

**Figure 4 | An illustration for the concept of flux share. a**, External magnetic flux can be decomposed into a uniform flux (penetrating through vortices, represented by Bean's model) and the rest (completely shielded, described by the London model). **b**, A type-II superconductor in magnetic field is analogous to a pebble in a stream. A part of flow penetrates through holes and the rest is completely detouring. **c**, Field profile within an infinitely long (macroscopic) cylinder is shown schematically by using the concept of flux share. **d**, The reversible magnetization shown in Fig. 3b can be interpreted as the Meissner share. A sum of the Meissner and vortices share is the total external magnetic flux.

Now, it seems to be natural to separate external field into two. The amount of external field completely shielded by the Meissner current will be defined from now on as 'Meissner share', $H_M$, and the rest, passing through vortices, as 'vortices share', $H_V$. As the field is increased, at first, the entire external field is the Meissner share, as shown in Fig. 4c and 4d. Just above the lower critical field, the vortices share will appear and then gradually increase. Near the upper critical field, most of the external field will be the vortices share. In Fig. 4c, the field profile along the diameter of an infinitely long cylinder is presented, as it is a widely used diagram for the explanation of Bean's critical state model [8]. In our previous argument, in Fig. 3d, the infinitely long cylinder is a microscopic object, size of vortex unit cell. But in here, it is a macroscopic object so that the field profile is uniform as shown in Fig. 4c as solid colored region. The exponential field decay near the boundary is presented as an additional line, which is obviously pretty much exaggerated. How we can decide the shares? In this simple un-pinned case, reversible magnetization is solely determined by the Meissner share and therefore the Meissner share is just equal to the reversible magnetization. The field dependence of reversible magnetization shown in Fig. 3b is the same as that of the Meissner share as shown in Fig. 4d. Using this share, the magnetization of an arbitrary shaped superconductor above the lower critical field can be readily simulated and that's the reason why the reversible magnetization or the Meissner share is approximated into a set of linear functions. Some of simulation results will be discussed in the next section.

**Reversible magnetization angular dependence**

Before we move on to a pinned case, we would like to discuss briefly on demagnetization effect of type II superconductors. Demagnetization effect can be understood as due to an internal field ($H_I$) enhancement with respect to the applied field ($H_a$) which can be describe by demagnetization factors, $D_j$, as $H_{Ij} = H_{aj}(1 + 4\pi D_j \chi)^{-1}$, where j denotes direction and $\chi$ is the magnetic susceptibility [13]. For a magnetic material under oblique field, the above internal field directional enhancement strongly dependent on the magnitude of magnetization and therefore, magnetization orientation varies as the field is increased. Usually, even for ferromagnetic materials, magnetization at very high field is quite low so that the demagnetization effect diminishes and the magnetization gradually aligns toward the field direction. However, if the magnetic flux is separable for type II superconductors as discussed in the previous section, there seems to be no reason for the magnetization orientation variation. A simulation result for a thin strip, size of 10 x 1.0 μm$^2$, under oblique field when it is tilted 60 degree with respect to the applied field is presented in Fig. 5a. The same penetration depth of 100 nm is used for the simulation and as noted, the Meissner share is the simplified set of linear functions shown in Fig. 4d. As the field is increased, the Meissner share is reduced but still that portion of external field is completely shielded. As a result, the reversible magnetization orientation does not change as shown as solid lines in Fig. 5e.

On the other hand, empirical results are quite different. An empirical result for a rectangular Nb bulk of 0.1 x 0.1 x 0.01 cm$^3$ in size at 5 K under oblique field is presented in Fig. 5d. The orientation of reversible magnetization starts to vary at a relatively low field, gradually aligns along the field direction, and then suddenly moves toward the shortest dimension of the sample. When the sample tilt angle is high, at 80 degree, the overall variation in the orientation angle is noticeably smoother. Early theoretical discussion over this matter is mainly focused on a case when the field is aligned along the principal axis of a superconductor [13,14]. It was shown that fundamental physical parameters, such as the thermodynamic critical field, $H_c$, are not affected by the demagnetization effect. The thermodynamic critical field is defined from a free energy difference between the normal and superconducting states, which can be obtained from the following integral relation [13,14],

$$\int_0^{H_{c2}} M \cdot dH = H_c^2/8\pi \,.$$

As can be seen in Fig. 5f, it is found that the thermodynamic critical field at various field orientations does not coincide with each other, contrary to the report of Hellman *et al*. [3]. For the applied field tilted at an angle of 30 degrees, the thermodynamic critical field is 1481 Oe, which is quite close to the reported value of 1480 Oe for Nb. But at 80 degrees, the estimated $H_c$ is only about 1368 Oe.

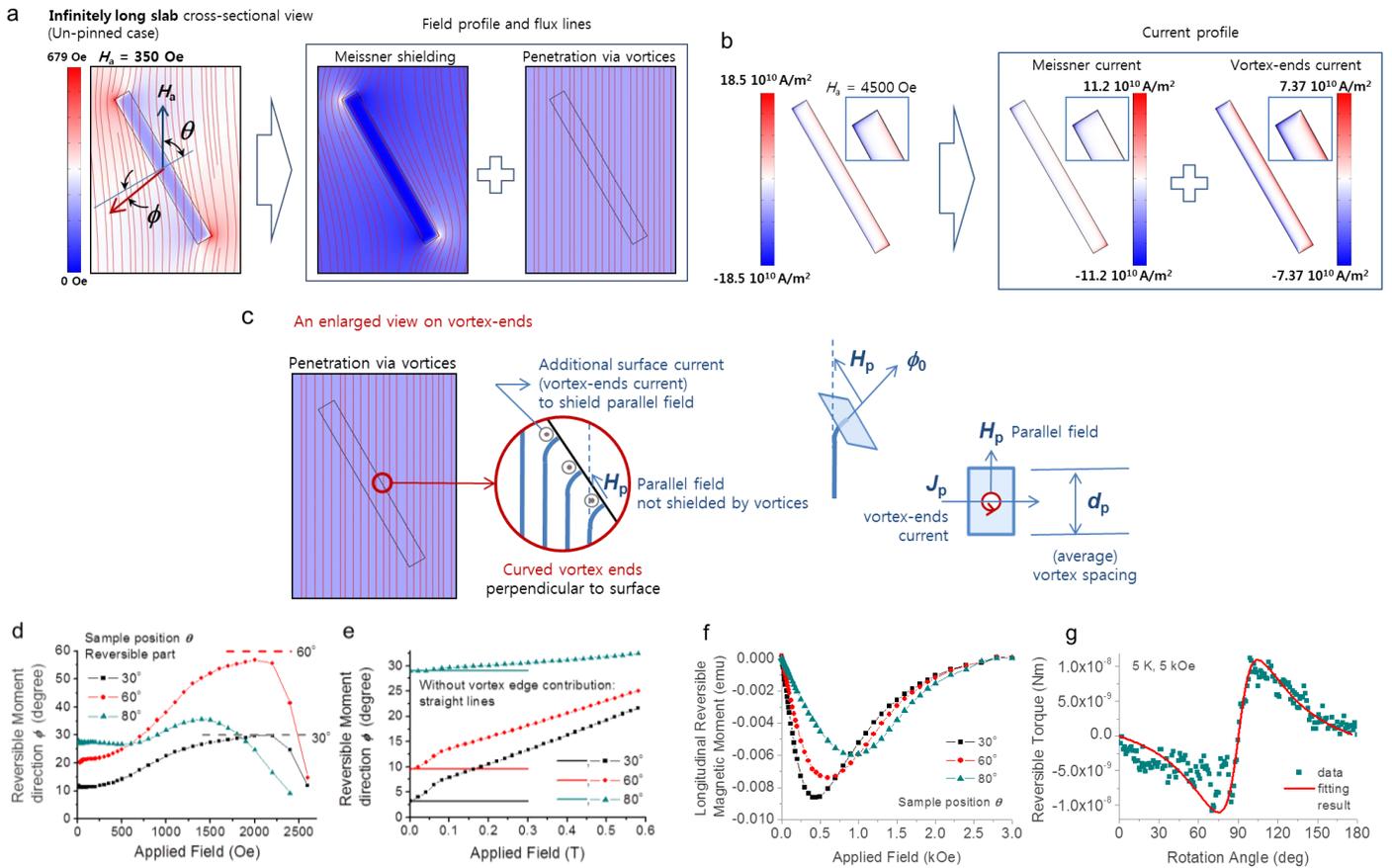

**Figure 5 | Reversible magnetization angular dependence and vortex-ends current. a**, Simulated field profile for an infinitely long slab. Superposition principle can be applied. **b**, Simulated current profile. Even though, there is no vortex density gradient, near the surface, there is 'vortex-ends' current due to vortex-ends curvature. **c**, An enlarged schematic view near the surface. From the consideration on the boundary condition, the amount of the vortex-ends current can be estimated. **d**, Empirical reversible magnetization angle variation as a function of field. Measurements have been carried out for a 0.1 x 0.1 x 0.01 cm$^3$ Nb bulk at 5 K. **e**, Simulated reversible magnetization angle variation as a function of field, with and without the vortex-ends current. **f**, Reversible magnetization along the field direction for the Nb sample at 5 K. **g**, Reversible part of an torque measurement for the Nb sample at 5K, 5 kOe. The torque data can be fitted by the anisotropic Ginzburg-Landau theory (solid line), with the anisotropy parameter, $\gamma$ of 4.5.

A plausible answer to these awkward behaviors has already been discussed by Hocquet *et al.* [15]. It was shown that vortices terminate perpendicular to the sample surface as schematically depicted in Fig. 5c. As discussed by Brandt, curvature of vortices means that there is a current proportional to its curvature [7]. This additional supercurrent will be referred as 'vortex-ends current', hereafter. The reason why the vortex-ends current might be a possible answer to the above two questions, why the magnetization angle varies and why the thermodynamic critical field is reduced at high tilt angle, can be guessed from Fig. 5b. In Fig. 5a, the field profile is presented as a sum of flux, perfectly shielded by the Meissner current and fully penetrated through vortices. Likewise, in Fig. 5b, the current profile is decomposed into Meissner shielding current and vortices contributions. Typically, there should be no vortices contribution as there is no pinning, but here, we need to consider the effect of vortex bending near the surface. Since the vortex-ends current is proportional to its curvature, the vortex-ends current will be more or less the same on most of the surface and therefore the magnetization by this current will be directed along the field direction. Also if the magnitude of vortex-ends current is quite substantial comparable to the Meissner shielding current, then it can affect the free energy and as a result, the thermodynamic critical field as well.

Hocquet *et al.* [15] also estimate the magnitude of the vortex-ends current, $j_{S0}$, and it was shown that $j_{S0}$ is proportional to the lower critical field and its curvature as $j_{S0} \propto H_{c1} \nu \times u / R$, where $\nu$ is the vortex directional and $u$ is the surface normal vectors, and $R$ the radius of curvature. In this work, a better estimation can be provided as the boundary condition is fully separated as shown in Fig. 5b. From the estimation of parallel field on the surface, the vortex-ends current can be calculated. If we adapt the vortex unit cell concept discussed in the previous section, the magnetic flux of a unit cell at the boundary can be described as a sum of flux by all neighboring vortices along the external field direction and by a central vortex, of which direction in this case, along the surface normal. Therefore, the parallel magnetic field, $H_p$, shown in Fig. 5c, right side, is proportional to the magnetic flux normal to the surface by an individual vortex, $\phi_n$. For convenience, we will define a fraction, $f = h/H_{c2}$, as a ratio of the local field ($h$) on the surface to the upper critical field. Then it can be argued that as the fraction approaches 1, as the local field on the surface approaches the upper critical field, the average magnetic field by an individual vortex is increased up to the lower critical field, $H_{c1}$. But as the local field on the surface approaches the upper critical field, the effect of vortices overlap becomes prominent. In the Hao-Clem model [10], the effect of vortices overlap was parametrized by a variational parameters, $f_\infty$, and it was shown that for high $\kappa$, $\kappa > 10$, $f_\infty$ can be approximated as [10],

$$f_\infty^2 = 1 - (h/H_{c2})^4.$$

Since the current density is proportional to $f_\infty^2$, it can be argued that the parallel field is proportional to $H_{c1} f(1-f^4)$. As the surface is tilted, it should be multiplied by $\sin \theta$. The unit cell area is elongated along the current normal direction, as shown in Fig. 5c, by $d_p$ which is proportional to $f^{1/2}/\cos \theta$, so that it should be divided by $f^{1/2}/\cos \theta$. Therefore, the estimated parallel field can be written as,

$$H_p = H_{c1} f^{1/2} (1 - f^4) \sin \theta \cos \theta.$$

If we include the vortex-ends current, general aspect of the magnetization orientation variation can be explained, as can be seen in Fig. 5e. As shown in Fig. 5b, the Meissner shielding current is more or less localized at the corners whereas the vortex-ends current is uniformly distributed on the surface. Since the vortex-ends current is proportional to $\sin \theta \cos \theta$, there will be no vortex-ends current when vortices are parallel or perpendicular to the boundary and the maximum will occur at 45 degree. As the vortex-ends current is increased, the magnetic moment due to this current gradually aligns the magnetization along the field direction. But still, the sudden movement toward the shortest dimension of the sample at very high field cannot be properly simulated. It needs to be noted that even though this additional surface current, the vortex-ends current is quite substantial but this current effect is not included in the free energy calculation. A more thorough study on the shape effect for a type II superconductor seems to be necessary. Another interesting empirical result can be found in Fig. 5g, the reversible part of torque data for the Nb sample at 5 K under the external field of 5 kOe. Though Nb is an isotropic superconductor, a typical result for an anisotropic superconductor can be observed, even can be well fitted by the anisotropic Ginzburg-Landau theory [16]. If reversible magnetization can be affected by irreversible magnetization than the observed weird torque result might be explained as a replica of the irreversible magnetization shape effect. The possibility of interaction between the two will be explored in the next section.

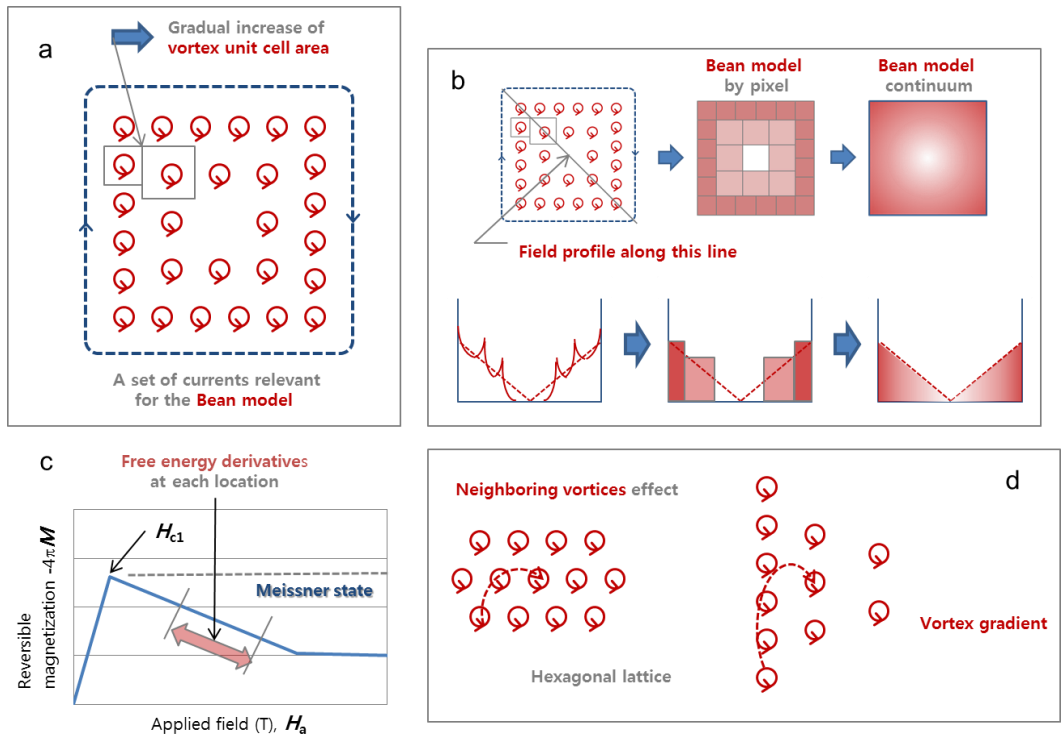

**Figure 6 | Bean's critical state model for a pinned case and local free energy derivatives. a**, Set of current relevant for Bean's model. **b**, A step-by-step process toward Bean's critical state model. **c**, If there is pinning, the vortex unit cell area depends on location and so does the Ginzburg-Landau free energy derivative, which can be estimated from reversible magnetization. **d**, The effect of vortex density gradient is somewhat similar to that of lattice structure difference. For example, if there is vortex gradient for a square lattice, it looks like locally a bit similar to a hexagonal lattice.

**Magnetization hysteresis**

In the first section, a simple un-pinned case for an infinitely long superconducting beam with a square cross-section has been discussed. It was shown that the entire currents inside can be grouped together, relevant for the London and Bean's critical state models as schematically presented in Fig. 1. If there is pinning, vortex density gradient is developed within a superconductor. Compared to Fig. 1, the only difference will be the set of current relevant for Bean's critical state model, as shown in Fig. 6a. The microscopic validity of Bean's critical state model has been argued by using the 'vortex unit cell' concept, shown in Fig. 3c. As we can use the same vortex unit cell concept whether the unit cell is located far away from the surface or not, this unit cell can be understood as a sort of a building block for Bean's critical state model. Can we extend this idea for pinned cases as well? As shown in Fig. 6b, each unit cell is converted into the corresponding building block for Bean's critical state model and then as we are mainly interested in macroscopic behaviors, each building block is approximated into a continuum object. The only difference compared to the un-pinned case is that the size of the vortex unit cell depends on position.

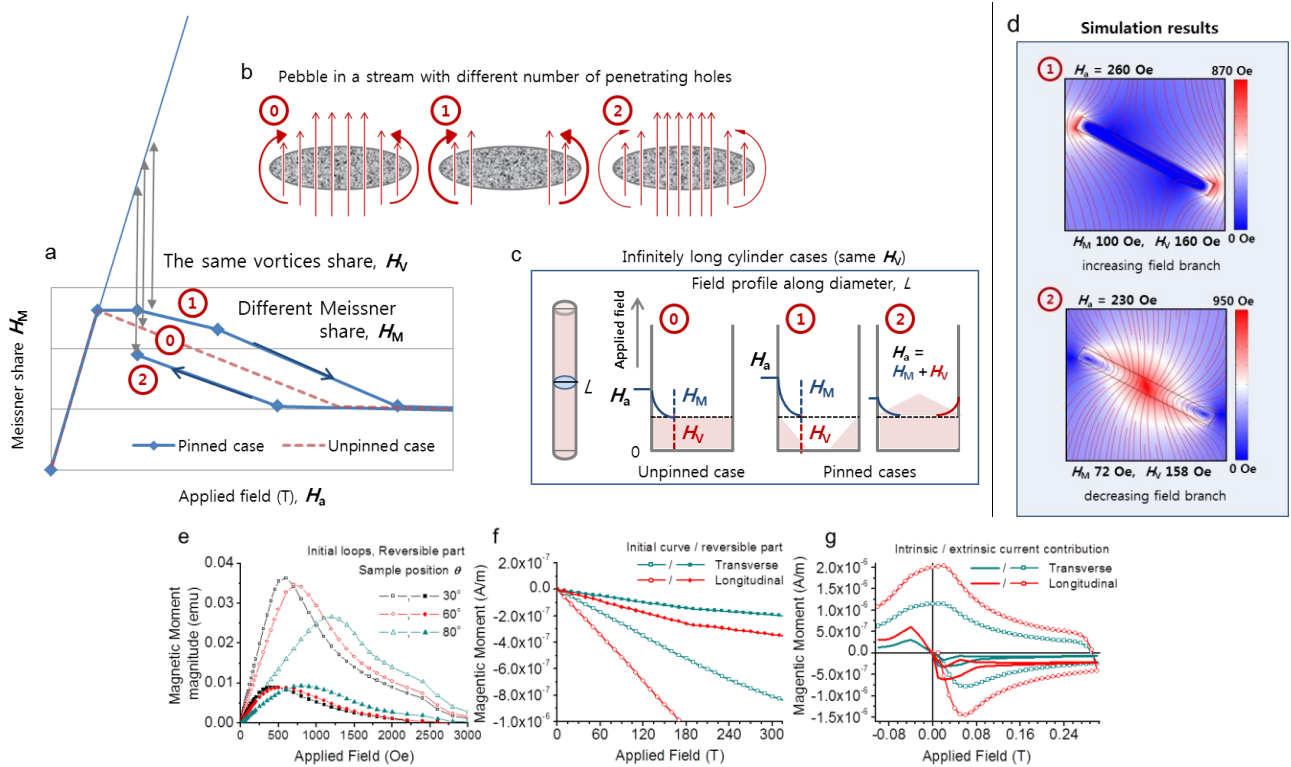

**Figure 7 | Irreversible flux shares for a pinned case. a**, The Meissner share is mainly determined by the number of vortices and therefore it is higher for the increasing field branch than that for the decreasing field branch. **b**, Pebble analogy. Detouring water-flow depends on the number of penetrating holes. **c**, The field profile within an infinitely long (macroscopic) cylinder is shown schematically for the increasing and decreasing field branches. **d**, Simulated field profiles for the increasing and decreasing field branches. **e**, Initial and reversible magnetic moment magnitude for the rectangular Nb bulk at 5 K. **f, g**, FEM simulation results for a superconducting strip. **f**, Longitudinal and transverse magnetic moments. **g**, Intrinsic and extrinsic magnetic moments. Intrinsic magnetic moment is not reversible.

Regardless the way of interpretation, Bean's critical state model has been a macroscopically acceptable model for decades for a variety of applications. The above interpretation is meaningful only when it can provide further information. In Fig. 4, for an un-pinned case, the idea of building block for Bean's critical state model naturally leads us to the concept of flux shares. What can be argued on the flux shares for a pinned case? One of the merits by using the concept of 'vortex unit cell' is that the free energy can be readily calculated in each cell because the effect of all neighboring vortices can be equivalently converted into that of a relevant reduced Bean current. The overall Ginzburg-Landau free energy inside a superconductor can be calculated and from its derivative, reversible magnetization can be obtained. We can do the derivative first. In each unit cell, the derivative of the Ginzburg-Landau free energy can be calculated, as shown in Fig. 6c. The free energy derivative at each location can be approximately estimated from the reversible magnetization curve. The average of the free energy derivatives will be the overall reversible magnetization. Since the free energy was calculated relative to the Meissner state and therefore whether we integrate first or differentiate first, results must be more or less the same. This kind of idea provide us a way to numerically simulate a superconductor when there is pinning. First, only Bean's critical state model is solved for a given geometry. The external field used in this simulation corresponds to the vortices share. At each vortices share, we can get a field profile within a superconductor from which the free energy derivative profile can be obtained and therefore the average, overall reversible magnetization or the Meissner share can be calculated. In other word, we can get a numerical relation between the external field and the vortices shares since the external field is just a sum of the Meissner and vortices shares. Using this relation, numerical simulation is carried out again. At this time, Not only Bean's critical state model but also the London model is solved, separately, and then the simulation results are summed up together to get overall results based on superposition principle.

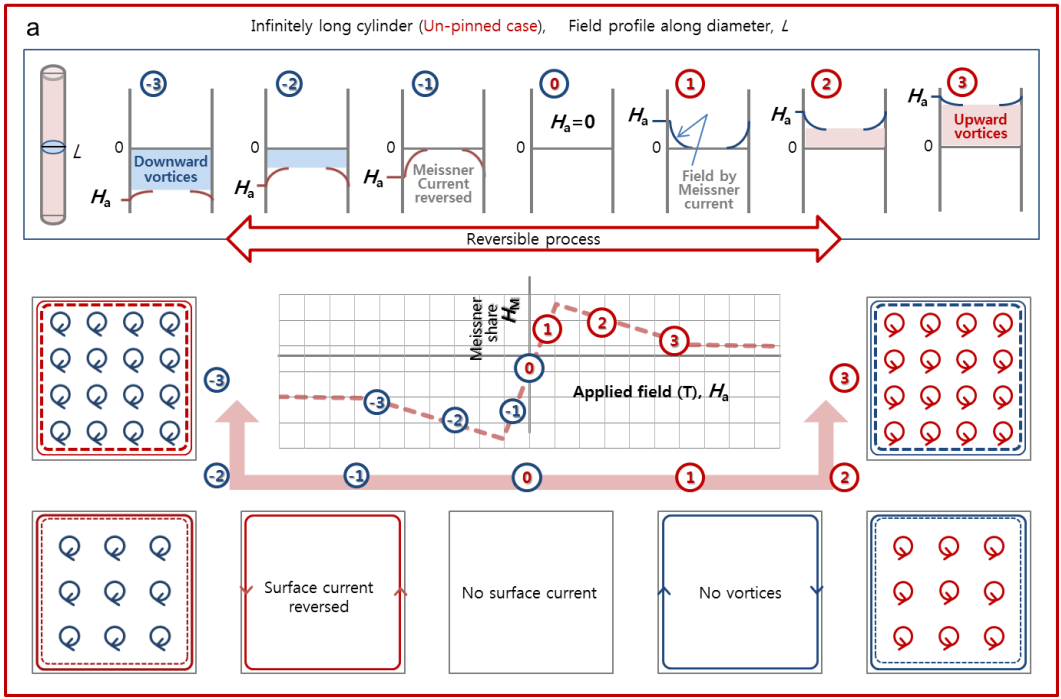

**Figure 8a | Flux shares, zero crossing process. a**, Reversible case. If there is no pinning, the way the surface current is reversed is quite trivial.

It should be noted that in the above discussion on the free energy calculation, the effect of neighboring vortices density gradient is not considered. Even though the free energy is dependent on the vortex array structure but the difference is typically less than 5% [8,10]. As shown in Fig. 6d, if there is a density gradient for a square vortex lattice, its shape is somewhat similar to a hexagonal vortex lattice. From this analogy, we would like to argue that the free energy will be affected by the effect of neighboring vortex density gradient but still mainly determined by the average vortex density at each location. Then, a very interesting consequence is that the Meissner share is irreversible as shown in Fig. 7a. Compared to an un-pinned case, small number of vortex will be in the superconductor on the increasing field branch and more vortices trapped on the decreasing field branch. In other word, the Meissner share is bigger for the increasing field branch and smaller for the decreasing branch compared to that of an un-pinned case, and therefore the flux shares are irreversible if there is pinning. It is similar to argue that the amount of flow goes around a pebble is mainly decided by the number of penetrating holes, as schematically depicted in Fig. 7b. The corresponding field profile diagram for an infinitely long cylinder case is presented in Fig. 7c.

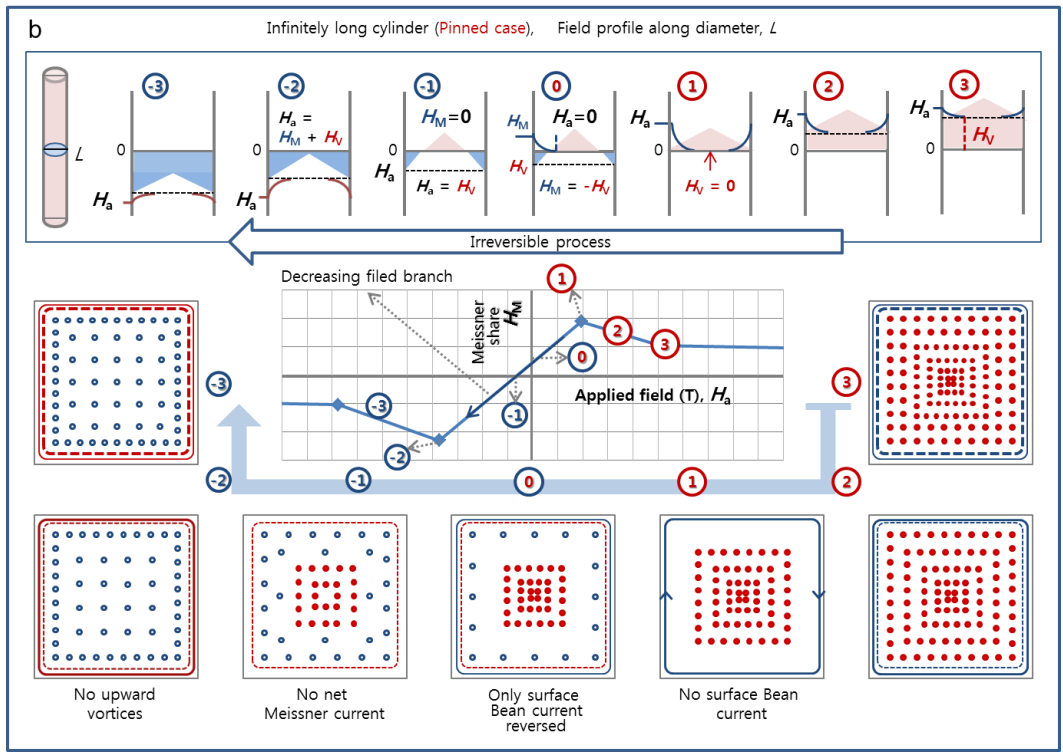

**Figure 8b | Flux shares, zero crossing process. b**, Irreversible case. As the field is lowered, first, the surface current relevant for Bean's model reduced to zero, at ①. Then the polarity of the vortices share is reversed. The applied field is zero, when the magnitude of the Meissner and vortices shares is the same, at ⓪. As the field is further lowered, the Meissner share changes its polarity at ⊖①, and then all vortices align downward at ⊖②.

Simulation results for a superconducting strip, size of 10 x 1.0 μm², a pinned case, are shown in Fig. 7d. An empirical power law relation, $E = E_0(J/J_c)^n$, has been used and the field dependence of the critical current density was assumed as, $J_c(B) = J_{c0}/(1+B/B_0)$. The parameters used are $10^{-5}$ V/m and 21, for $E_0$ and the $n$-value, 2 x $10^{11}$ A/m² and 300 Oe, for the $J_{c0}$ and $B_0$, respectively. As mentioned in the introduction, the magnetization slope, $-dM/dH$, is quite different depending on whether it is measured from an initial loop or from reversible magnetization, for example, as shown in Fig. 7e. Fig. 7e is measured data for the 0.1 mm thick Nb sample, 0.1 x 0.1 x 0.01 cm³ in size at 5 K. The above irreversible flux shares can explain the huge difference of the magnetization slope in a reasonable way. Simulation results for the superconducting strip tilted 30 degree with respect to applied field are shown in Fig. 7f. In accordance with a general argument that reversible and irreversible quantities are not interact with each other, the two sets of partial differential equations one for reversible and the other for irreversible magnetization are solved separately and then summed up together, based on superposition principle.

Neither demagnetization field effect of the trapped vortices nor the Meissner screening current interference with extrinsic current over a penetration depth is considered [3]. It may be argued that reversible and irreversible properties do not interact with each other directly but they are interconnected indirectly by the flux shares. Pinning prohibits the entrance or exit of vortices, changes the number of vortex inside and thereby affects the reversible magnetization as well. Therefore, if we still refer the magnetization by the surface Meissner current as 'reversible' magnetization then we can say that the 'reversible' magnetization can show hysteresis, as shown in Fig. 7g, which can be the reason for the anomalous torque measurement result discussed in the previous section. To avoid complication, from now on, the surface current relevant for reversible magnetization will be referred as 'intrinsic current' and all the currents relevant for Bean's critical state model, as 'extrinsic current'.

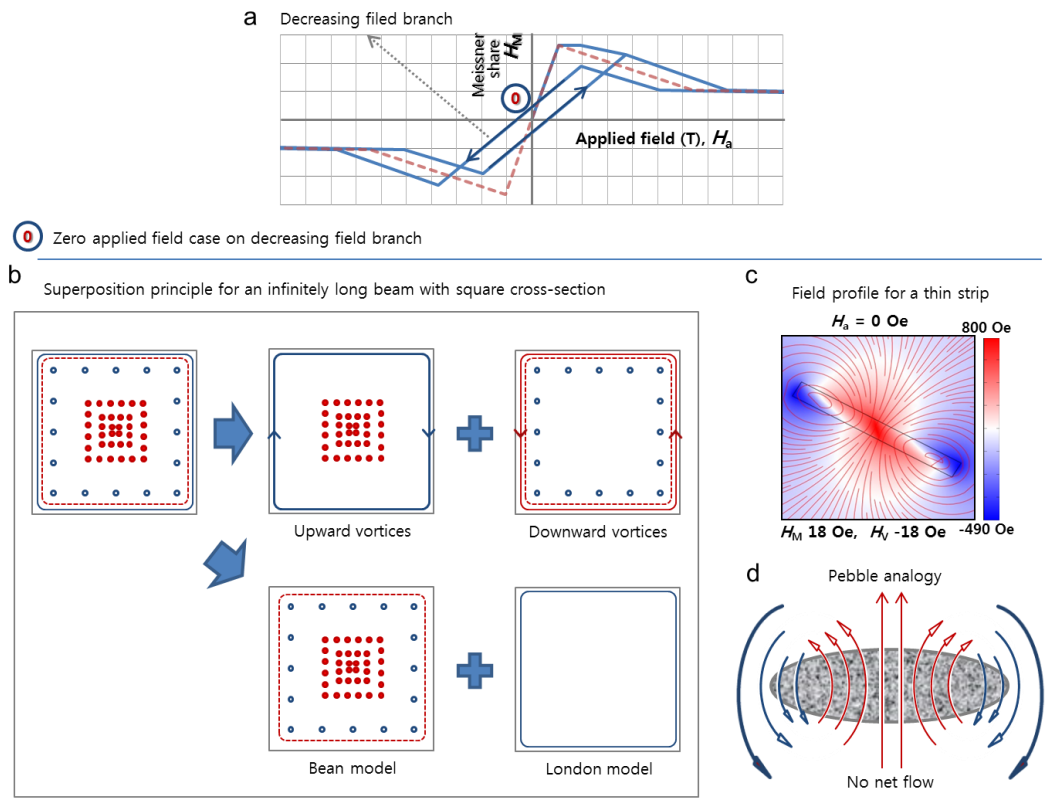

**Figure 9 | Flux shares at zero field. a**, Irreversible flux shares near zero field region. **b**, A zero field case can be decomposed as a mixture of up and downward vortices or as a sum of Bean's and the London model. **c**, Simulated field profile at zero field. **d**, Pebble analogy. Water-flow goes through the holes and around the boundary, so that there is no net flow.

Before we go on further, another interesting aspect of the irreversible flux shares needs to be discussed. As shown in Fig. 8a, if there is no pinning, as the field is lowered down to zero and then changed its polarity. The way how the surface current is reversed is quite trivial. But for a pinned case, as shown in Fig. 8b, it is not that simple. As the field is lowered from ③ to ①, the Meissner share increases as the number of vortices are reduced. The Meissner share reaches its maximum when the number of vortices along the positive field direction is in its minimum, just before the negative field directional vortices appear, when the vortices share is zero. If we further lower the field, vortices oriented along the negative field direction appear from the boundary and therefore the surface current relevant for Bean's model changes its direction. When vortices with opposite directions coexist within a sample, it is easier to consider the case as a combination of separate upward and downward vortices cases, as shown in Fig. 9. Now, we can calculate the flux shares for the upward and downward vortices cases separately and then total flux shares can be summed up. From this view point, it can be argued that, if there are trapped vortices, zero applied field for a superconductor is in fact a superposition of two opposite applied fields, the same Meissner and vortices shares with different polarities. This corresponds to a case where water-flow through central holes circles around outer holes and surface so that there is no net flow as shown in Fig. 9d. As the field is further lowered, at ① in Fig. 8b, the number of upward and downward vortices is equal to each other and at this point, the corresponding Meissner share is the same but different in their polarity and therefore no net Meissner share. And then, at ② in Fig. 8b, upward vortices disappear and the Meissner share reaches its local minimum.

**Paramagnetic Meissner effect**

Complete magnetic shielding regardless the way of cooling, Meissner effect, is the characteristic feature of superconductivity, which differentiate superconductivity from perfect conductivity. However, for type II superconductors, if field cooled, cooled from above the transition temperature under magnetic field, magnetic flux within the sample is partially remained inside in the form of quantized vortices. Even though the exact amount of trapped vortices is not yet clearly understood, it had been usually argued that magnetic flux density by the trapped vortices cannot exceed the external flux density so that a diamagnetic response is still expected. From about 20 years ago, on the other hand, plenty of empirical evidences have been reported that the opposite can occur, which usually called as paramagnetic Meissner effect [6,17-20]. It was found that the paramagnetic response becomes more evident as the field is lowered, when the sample is aligned perpendicular to the field and the surface is relatively dirty [6,18]. The origin of this peculiar phenomenon has been argued as related with exotic intrinsic reasons [19], such as $\pi$-junctions or $d$-wave mechanisms because at first, most of reports were on cuprate superconductors [6,17]. It became more complicated as conventional type II superconductors, such as Nb or Al, also reported to show paramagnetic Meissner effect [6,18]. Extrinsic and less exotic reasons, such as flux compression [20], whether it is originated due to sample inhomogeneity or by temperature gradient, have been argued. But still, the origin of paramagnetic Meissner effect, why it is affected by the field magnitude, sample orientation and surface cleanness are not clear yet.

Here, we start with a question how we can assess the vortices trapped inside when a superconducting sample is field cooled and then move forward to its consequences. As we lower the temperature from above the transition temperature, the superconductor will pass through from a normal state, the upper and lower critical fields, finally be in a mixed state. The situation is quite similar to the upper branch of a magnetization hysteresis measurement as schematically shown in Fig. 10a. Temperature dependent magnetization curves are simply represented as a set of linear lines as shown in the left side and each magnetization curve is separately presented in the right side as the temperature is lowered by using arbitrarily scaled units. At first, a case when the temperature is just lowered down below the transition temperature corresponds to near the upper critical field region and then as the temperature is further lowered, corresponds to a near zero field case in the magnetization hysteresis curve. From this analogy, we would like to argue that not only the external field but also the temperature variation can affect the amount of vortices trapped inside and thereby the flux shares as well. Trapped vortices are, in some sense, just partial gateways for the external magnetic flux and only a reduced amount of the external field is shield by the intrinsic current.

At first, when the temperature is lowered just below the transition temperature, vortices are generated inside and most of the magnetic flux will pass through vortices so that the Meissner share will be quite negligible.

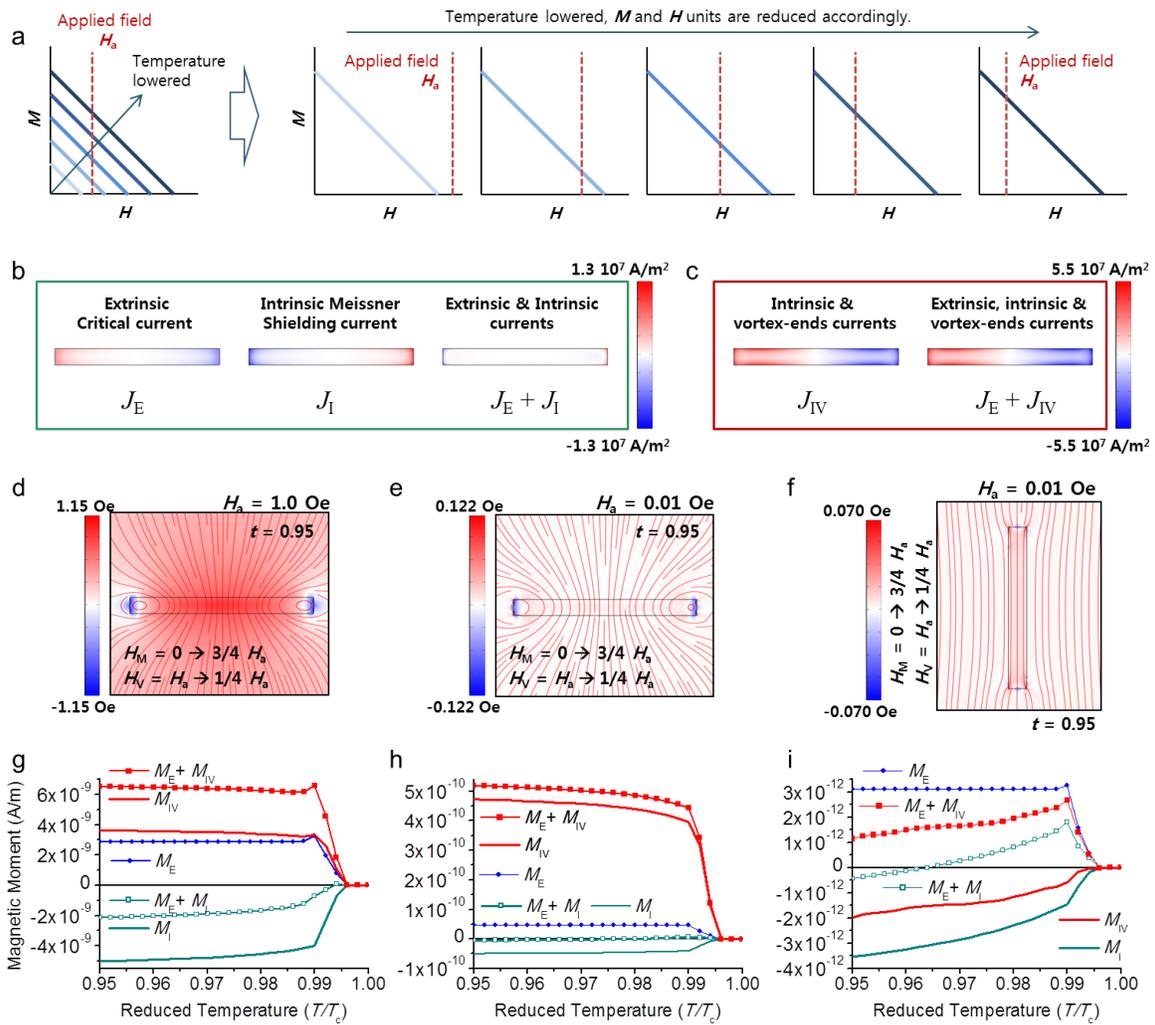

**Figure 10 | Origin of the paramagnetic Meissner effect. a**, Similarity between field-cooled magnetization and the upper branch of a magnetization hysteresis loop. **b, c**, Current profiles without and with vortex-ends current within a strip during a cool-down process for a perpendicular field of 0.01 Oe. **d, e, f**, Magnetic flux density profiles for a strip, perpendicular and parallel to applied fields of 1.0 and 0.01 Oe. **g, h, i**, Magnetic moments for each case. Magnetic moments by the current defined at **b, c**, are separately presented. Total magnetic moment (solid squares) becomes more paramagnetic as the field is lowered and aligned in perpendicular direction.

As the temperature is further lowered, however, a portion of vortices is expelled from the surface and therefore the Meissner share will be substantially increased. Even though the applied field remains the same, the Meissner share is increased whereas the vortices share is reduced as the temperature is lowered. As a result of this reduction in the vortices share, vortices are not just trapped inside but vortex density gradient or curvature is developed inside. In other word, to compensate this field variation, extrinsic critical current flows within a superconductor in opposite direction with respect to the intrinsic supercurrent. The generation of extrinsic critical current in opposite direction is similar to the idea of flux compression [20] but here the extrinsic current is originated not by sample inhomogeneity or temperature gradient but due to the flux share variation.

Is this opposite directional critical current good enough to be the reason for the paramagnetic Meissner effect? In order to simulate a field cooled case, all we need to know is how the flux shares vary as the temperature is lowered. Temperature dependence of other superconducting parameters can be found in literature. For example, the thermodynamic critical field can be written as a function of reduced temperature [16], $t = T/T_c$, as, $B_c(T) = B_c(0) (1 - t^2)$. Empirically, magnetization drops rapidly almost just immediately after the temperature is lowered below the critical temperature and then saturates [17,18]. Therefore, the Meissner share is assumed to be zero till the reduced temperature, $t$, is lowered down to 0.995, increases linearly, rapidly till $t$ reaches 0.99 up to 3/4 of the applied field and then saturates. Simulated current profiles for the strip are shown in Fig. 10b. Detailed simulation parameters can be found in Supplementary Information. As shown in Fig. 10b, the extrinsic critical current during the cool-down process flows in opposite direction but is still smaller than the intrinsic supercurrent, not good enough to explain the paramagnetic Meissner effect. In this simulation, we assume that the Meissner share when saturated is only 3/4 of the applied field. As discussed in the previous section, the flux shares near zero fields can be quite complicated. The Meissner and vortices shares even can have different polarities. The saturation Meissner share can be much higher than 3/4 of the applied field. But this does not change the situation much. Even if we increase the saturation Meissner share, the extrinsic current is still smaller than the intrinsic supercurrent. However, if we included the vortex-ends current, which was discussed in the second section, the opposite directional current is significantly enhanced, surpassing the intrinsic current as can be seen in Fig. 10c, good enough to explain the paramagnetic Meissner effect.

Further simulation results, field profile, flux line distribution and temperature dependence of magnetization for three different cases are presented together in Fig. 10. Only the flux lines produced by the vortices are shown in Fig. 10d, 10e, 10f, to clarify their impact on the vortex-ends current. One of the characteristic features of the paramagnetic Meissner effect, the sample orientation dependence, can now be argue as related with this flux line distortion. As discussed in the second section, the vortex-ends current is proportional to $\sin\theta\cos\theta$ so that it is more prominent when the vortex line are angled to the surface. If we compared the two perpendicular cases shown in Fig. 10d and 10e with the parallel field case, Fig. 10f, the flux line distortion on the surface is much more severe for the perpendicular cases. We argued that this is the reason why the paramagnetic Meissner effect can be observed when the sample is align perpendicular to the field, since more vortex-ends current is induced. In Fig. 10g, 10h, and 10i, the magnetization by the extrinsic, $M_E$, intrinsic currents, $M_I$ and intrinsic currents including the vortex edge current, $M_{IV}$ are presented together with overall sum, $M_E + M_I$ and $M_E + M_{IV}$. The magnetization by the vortex-ends current is not only substantial but also paramagnetic. Furthermore, the paramagnetic response is enhanced as the field is lowered, as can be seen from Fig. 10g and 10h. The reason for this peculiar field dependence is because, as discussed, the vortex-ends current is proportional to the square root of the field. When the field is reduced from 1 to 0.01 Oe, for example, all the other magnetizations are reduced to 1/100 but the paramagnetic magnetization due to the vortex-ends current only lowered by 1/10. The impact of surface polishing, which is known to reduce the paramagnetic Meissner effect, also can be understood. Surface polishing generates Bean-Livingston barrier [21], which facilitates vortices movement toward sample outside and thereby reduce the number of trapped vortices. This will reduce the vortex-ends current and thereby the paramagnetic Meissner effect as well.

In summary, a macroscopic model for the description of both the reversible and irreversible properties of type II superconductors has been proposed. The main new idea is the concept of flux shares. A part of external magnetic flux is completely shielded while the rest penetrates through vortices. By using the concept of flux shares, numerical simulations can be carried out, by solving the London and Bean's model separately and then sum up together based on superposition principle. It was shown that the flux shares are irreversible, which means even though reversible and irreversible properties are separable but can affect each other. Near the boundary, vortices are curved, orthogonal to the surface and the current at that curvature is specifically named as vortex-ends current. It was shown that the vortex-ends current is quite critical for the characteristic features observed in the angular dependence of reversible magnetization and most notably for the paramagnetic Meissner effect. Further detailed studies are anticipated. For a quantitative analysis, the free energy calculation including vortex-ends current needs to be carried out first. Also, as the intrinsic and extrinsic properties are link together, shape effects need to be carefully included. The macroscopic electromagnetic description of type II superconductors can be useful for a variety of fields. Precise determination of fundamental physical parameters from magnetization data, a better AC loss estimation may be possible.

**Acknowledgements**

We would like to thank Mahn-Soo Choi, Jae Hyuk Choi, Kisung Kwak, In Sik Choi and Kwang-il You for their helpful discussions and comments. This work was partially supported by Mid-career Researcher Program through NRF grant funded by the Ministry of Education, Science & Technology (MEST) (No. 2010-0029136)**.**

# Beyond Bean's critical state model: On the origin of paramagnetic Meissner effect

Sangjun Oh, Dong Keun Oh, Won Nam Kang, Jung Ho Kim, Shi Xue Dou, Dojun Youm and Dong Ho Kim

# SUPPLEMENTARY INFORMATION

**METHOD SUMMARY**

Nb bulk samples were cut from a commercial 99.95% sputtering target. Magnetic moment and torque measurements have been carried out with Quantum Design Magnetic Property Measurement System (MPMS) and Physical Property Measurement System (PPMS). FEM simulations are carried out with Comsol Multiphysics. Details of simulation methods are discussed below.

**Supplementary Information I: Microscopic basis for Bean's critical state model**

**Effect of neighboring vortices and that of corresponding reduced surface current**

One of the major argument points of this work is that the effect of (all) neighboring vortices can be equivalently described by corresponding reduced surface current, which means that a vortex unit cell can be treated by the same way whether it is located near the surface or not. Therefore, we can treat the vortex unit cell as a sort of a building block for the Bean's critical state model. As a further evidence for this statement, we would like to show a bit more cases, as shown in Fig. S1. For a given average magnetic flux, **B**, for example for 10 and 100 Oe, vortex unit cell area (*A*), unit cell radius ($a_0$) were calculated and a hexagonal vortices array was built up. Firstly, the current profile by neighboring vortices was compared with that of the corresponding reduced surface current. The surface current was estimated from an assumption that it is equivalent to that of infinitely long cylinder of radius, *a*, placed in parallel field, **B**. This current is continuous whereas the vortices (and their currents) are discrete, and therefore, to valence the Lorentz force by the vortices and by the reduced surface current, the cylinder radius (*a*) needs to be a bit larger than the unit cell radius ($a_0$). For the hexagonal array, when the cylinder radius was about 1.6 times larger than the unit cell radius, the current profiles and therefore the Lorentz force are relevant with each other. (In the above figures, only the profiles within the unit cell areas are presented for a bit more clear comparison.) The field profiles generated by the above currents are also shown in Fig. S1. As the field is increased, the field overlap by neighboring vortices is quite severe as can be clearly seen Fig. 3d of the main manuscript. And similar to the overlap field effect, there can be found almost the same background field in the solution for an infinitely long cylinder superconductor. As can be seen as below, the solution for an infinitely long cylinder superconductor, proportional to the 0th order Bessel function, $I_0$, which saturates as the normalized radius ($\rho/\lambda$) approaches zero.

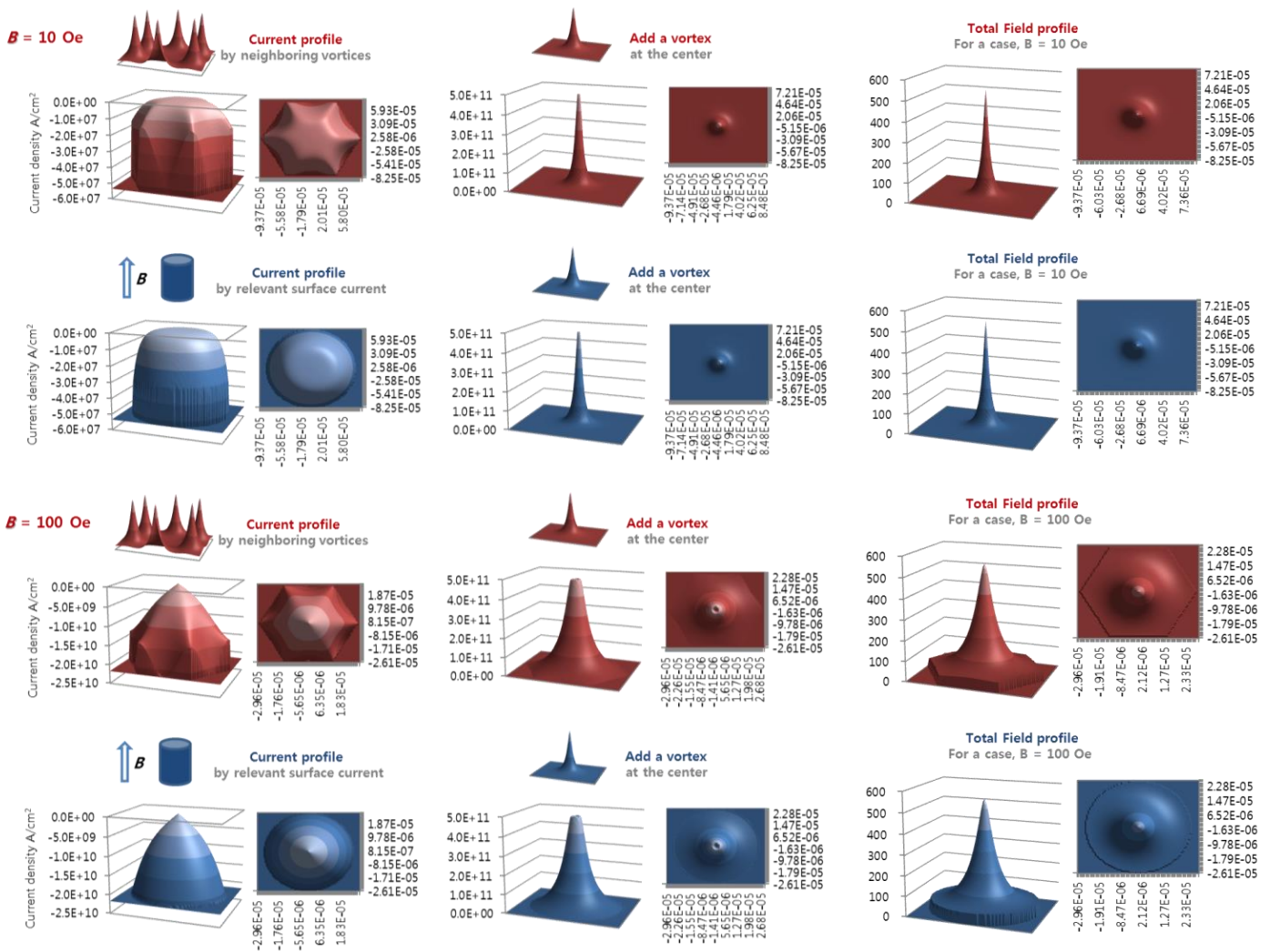

**Figure S1 | Effect of neighboring vortices and corresponding reduced surface current at other fields. The first and third rows:** From left to right, the current profile by neighboring vortices, the total current and field profiles including the central vortex within the vortex unit cell for $B$ = 10 and 100 Oe, respectively. **The second and fourth rows:** From left to right, the equivalent surface current profile, the total current and field profiles including the central vortex within the relevant unit cell circle, for $b_0$ of 10 and 100 Oe, respectively.

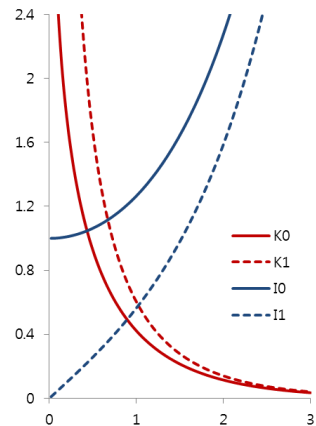

**Figure S2 | Modified Bessel functions.** For an infinitely long cylinder, the current is proportional to a modified Bessel function of 1st order ($I_1$), whereas the field is described by 0th order ($I_0$) as a function of normalized radius with respect to the penetration depth ($\lambda$), $\rho/\lambda$. As the field is increased, the normalized radius is lowered below 1 so that the field overlap by neighboring vortices can be proper simulated.

**Supplementary Information II: Reversible magnetization angular dependence**

**Angle dependent magnetic moment measurements**

Magnetic moments were measured by Quantum Design (QD) Magnetic Property Measurement System (MPMS) using a horizontal rotator. The SQUID system has both longitudinal and transverse pick-up coils, which are parallel and perpendicular to the applied field, respectively. The configuration of measurements is as schematically depicted below. The longitudinal and transverse magnetic moment measurement results for $MgB_2$ thin film, as an example, are also shown in Fig. S3. During sample mounting procedure, there can be a slight misalignment so that nominal angle can be different from actual angular position. For the $MgB_2$ thin film, the misalignment is about 7.8 degrees and was calibrated. The same calibration procedure has been carried out for the Nb disk sample.

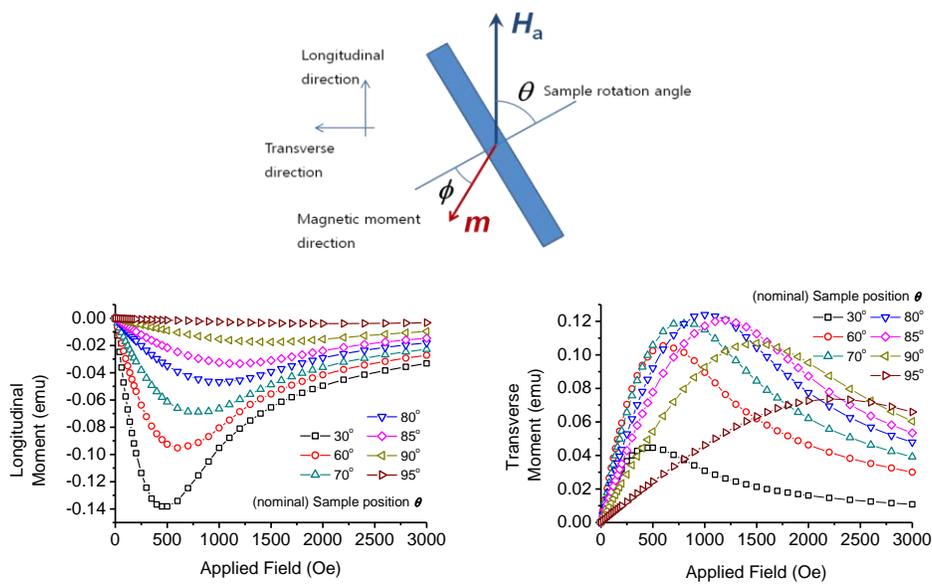

**Figure S3 | Magnetization Measurements. Top:** Measurement configuration. **Bottom:** Longitudinal (left) and transverse component of magnetic moment for 1 μm thick $MgB_2$ thin film at 5 K.

**Angle dependent reversible magnetization simulation**

London equations can be easily incorporated into commercial FEM software and therefore, if there is no pinning, the only problem left is how we can set a proper external field for the London model when total external field exceeds the lower critical field. A new concept of flux share was discussed in this work and it was argued the Meissner share, which is basically the same as the reversible magnetization for un-pinned case, is the only external field which needs to be solved by the London model. As shown in Fig. S4, the Meissner share is simplified as a set of linear functions to incorporate it easily into the FEM solver and further, shape effect is taken into account. The used parameters are the penetration depth of 0.1 μm, the lower and upper critical fields of 0.05 and 0.7 T, respectively.

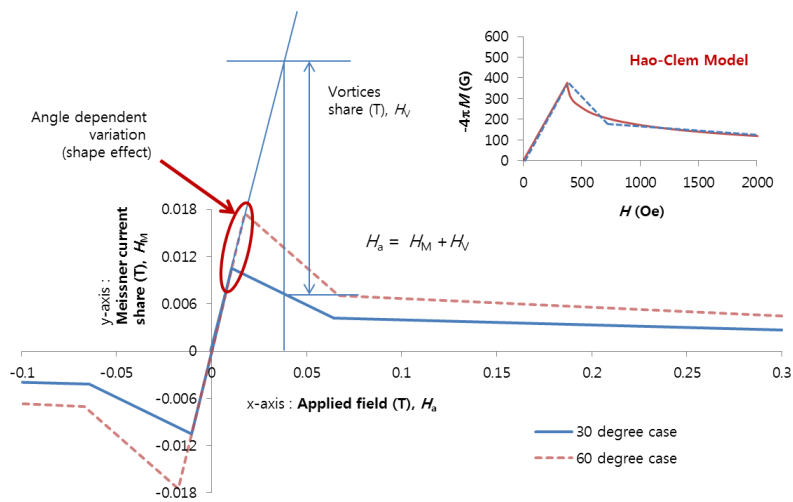

**Figure S4 | Proper external field for the London model, the Meissner share.** (**Inset:** For simplicity, the Meissner share is assumed as a set of linear functions.) For angular dependence simulation, shape effect needs to be considered. The demagnetization factor is directly obtained from a dummy simulation and the linearized Meissner share is adjusted accordingly.

**Supplementary Information III: Magnetization hysteresis**

**Irreversible Meissner share for pinned cases**

As discussed in the previous section, for an un-pinned case, a simplified set of linear functions (redrawn in Fig. S5a) is used for the simulation as the Meissner share. If there is pinning, this share will be affected by the number of vortex inside a superconductor. Since the average magnetic flux density ($B$) is proportional to the number of vortex, the number of vortex can be estimated from an average magnetic flux density. The average magnetic flux density for a give field can be estimated from a dummy simulation for Bean's critical state model, a conventional Bean model simulation without considering reversible magnetization. A dummy simulation result is shown in Fig. S5b. From this simulation result (Fig. S5c), the irreversible Meissner share has been deduced as shown in Fig. S5d.

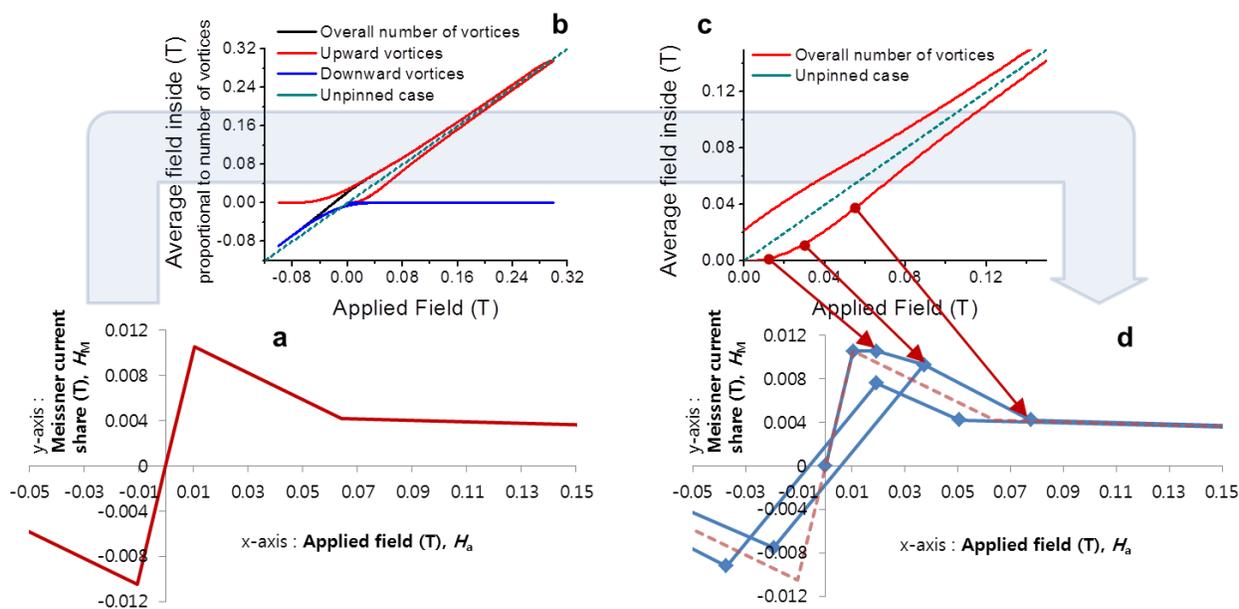

**Figure S5 | Determining irreversible Meissner share. a,** A reference reversible Meissner share. **b,** Average field inside the strip, which is proportional to the number of vortex, is obtained from a dummy simulation. **c,** The total number of vortex is further simplified linearly. **d,** Corresponding modification on the Meissner share decided from the overall number of vortex.

**Supplementary Information IV: Field cooled (FC) magnetization simulation**

**Simulation parameters: Temperature dependence**

The temperature dependence of the penetration depth can be written as, $\lambda(T) = \lambda(0)/\sqrt{1-t^4}$, where $t$ is the reduced temperature defined as, $t = T/T_c$. The penetration depth at zero temperature is assumed as 0.1 μm. The temperature dependence of the critical current density, $J_c(T)$ and the $n$-value is assumed as follows. The critical current density is tentatively assumed to be inversely proportional to the penetration depth as, $J_c(T) = J_c(0)\sqrt{1-t^4}$, where $J_c(T)$ is set to about 1/10 of the de-pairing current density. The field dependence of the $n$-value is recently reported [reference S1]. It was further assumed that the temperature dependence of the $n$-value is similar to its field dependence. The $n$-value is assumed as, $n(T) = n(0)\exp(-\exp(-60(1-(1-t)/0.97)))+1$, where $n(0)$ is set to 20. The temperature dependences of the simulation parameters are presented in Fig. S6.

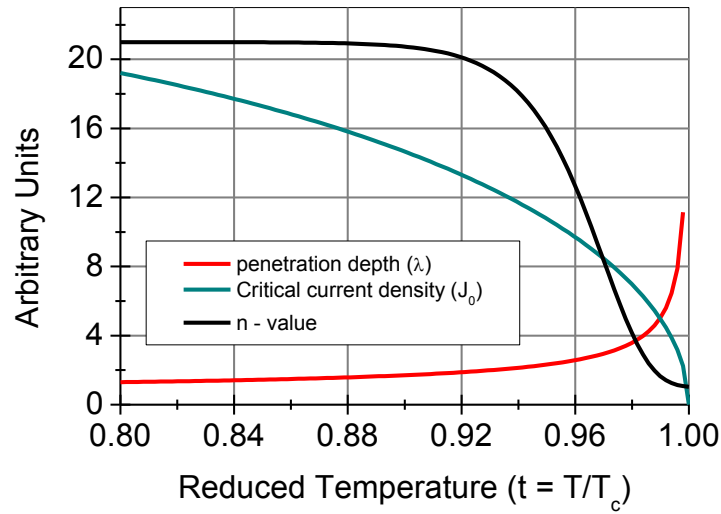

**Figure S6 | Temperature dependences of simulation parameters, $\lambda$, $J_c$ and the $n$-value.**